\documentstyle[aps,prd,epsfig]{revtex} 
\draft

\begin{document}

\title{
\hbox to\hsize{
\hfil }
\vskip 36pt
Presence On Earth of Non-Identical Particles}
\author{L.~Stodolsky}
\address{Max-Planck-Institut f\"ur Physik 
(Werner-Heisenberg-Institut),
F\"ohringer Ring 6, 80805 M\"unchen, Germany  
~~~~~~~~~~~~~~~~~~~~~~~~~~ email: les@mppmu.mpg.de}

\maketitle
\bigskip

\begin{abstract} We speculate on some novel
consequences possibly associated with a change of 
 elementary particle constants with
time.  These concern the question as to whether
 particles reaching the earth  from 
distant parts of the universe reflect their properties at their
point of origin  and so can be different from local particles. 
Assuming that such particles would be present in very high energy
cosmic rays, we estimate their density on the earth by two
different methods, finding on this hypothesis a density of roughly
$10^{-28}-10^{-29}$. Various possibilities for detection are
discussed,
including the formation of unusual nuclear isotopes.
\end{abstract}

\vskip2.0pc

 At least since Dirac's ``large numbers'' paper~\cite{dirac}
there has been speculation about a possible time dependence of the
fundamental physical constants, a fascinating idea that
has never   been convincingly confirmed or refuted.
The discussion 
 received new impetus with the claim~\cite{claim}  that 
observations of high
redshift spectral lines  indicate a value of $\alpha$  (the fine
structure constant)   differing from its accepted value
as determined  in the vicinity of the earth, by $ \sim 10^{-5}$.
This has   stimulated  various theoretical
speculations~\cite{theo}. 

Here we would like to add a further question raised by the
possibility of such phenomena. In  different regions of spacetime
there would 
presumably  be different parameters  for elementary
particles, and depending on how the variable constants are
implemented,~\cite{theo} we could have different charges for
elementary
particles, different electron/proton mass ratios, different
magnetic moments for the neutron, and so forth. If particles can
have different properties in different regions of spacetime, then
what happens when there is contact between these regions, as
through the agency of very high energy cosmic rays?
  
In general we might expect two kinds of situations. In one, the
values of
the constants would depend, say, on some kind of local field, which
then varies in spacetime. Thus
independent of any history,  local properties are always the same
(however see the remarks about adiabaticity below).
This  picture is as in general relativity, and is probably the most
likely one.

 However there is another possibility  that we would like to
consider briefly. This is that particles traveling from one region
to another retain  to some degree or another their properties  and
so are different than  local particles. Without a detailed theory
it is of course difficult to analyze this question fully, but it is
an interesting one to speculate upon, particularly from the point
of view of detection.

 In this later picture  the properties of a particle could depend
on
its history. For example, as seems plausible in such theories, the
{\it rate} of
change of physical constants and not the constants themselves
could depend on local conditions. In this case the properties of a
particle 
would depend on the space-time path it used to reach us. In fact
 Weyl's original ``gauge invariance'' had this property, and  
was rejected by Einstein for this very reason; an atom's
 spectral lines would depend on its history~\cite{weyl}.
We know from everyday experience that  such history-dependent
effects cannot be very large or very frequent. But it is perhaps
not
entirely excluded that they are present very weakly or very rarely.
While such a situation would open a Pandora's box of difficulties
with respect to such basic principles as gauge invariance and the
exclusion principle, it is perhaps amusing to briefly consider a
few of the consequences. 

Very high energy
cosmic rays would appear to particularly interesting for this
question. These probably reach the earth from distant parts of
the
universe. Furthermore it is worth  noting that   they do so
in a relatively short  time, measured in terms of their own proper
time.

  For energies above the
``knee'' at $\sim 10^{16} Ev $ in the cosmic ray spectrum, protons
and nuclei cannot
be contained by the magnetic field of the  Milky Way, and similar
behavior should obtain for other galaxies. In
intergalactic space they travel essentially
freely and at these energies there is nothing to stop them coming 
to the earth
from great distances. Hence it is 
conceivable that  presently on the earth  there are a small number
of ``alien''~\cite{harald} particles --call then ``alieons'' --of
remote extragalactic origin.
 
In addition to their distant origin,
a second point is the high velocity of the cosmic rays.  If  
 the proper time for the particle plays a role and  fast
relativistic
particles ``age'' more slowly,  then ``alieons'' originating from
high energy cosmic rays
are very ``young''. The time dilatation factor for  a
$10^{16} Ev$
proton is $\gamma \approx 10^{7}$ so that a high energy cosmic
proton
 which has been  travelling during the whole lifetime of the
universe, would upon
 reaching us, be only
about a thousand years ``old''. Naively, we
might then expect that  its properties have ``aged'' very little
compared to those for a local  proton. Of course, without a
detailed theory we do not know how the ``ageing'' process takes
place nor if the time dilatation effect is not compensated by
perhaps some tensor property of the fields involved. However,  this
would seem very curious, ---in an admittedly already very curious
situation-- and underlines the need to understand the properties of
transport in such theories.
The particles in questions are essentially
nucleons since protons and nuclei are the dominant
component of the high energy cosmic rays. Alieon neutrons from 
incoming  nuclei may be of special
interest 
since after slowing down they can be captured and remain in 
ordinary nuclei.

 On the practical level, the important issue is of
course the number of such particles present.
One way to guess the number of extragalactic particles on the
earth would be to simply take all particles above the ``knee'' as
representative of the incoming extragalactic flux. 
The rate of cosmic rays for the region of the ``knee'' is about one
per square meter per year. Multiplying  by the area of the earth's
surface and for a time period of $1.2\cdot 10^{10}$ years we obtain
$\approx
6 \cdot 10^{24}$ 
for the number of accumulated extragalactic particles,
corresponding to   about 0.01 grams for protons or 0.5 grams
for nuclei. This
is to be compared
with $6 \cdot 10^{27} g $ for the mass of the earth. Hence the
density of
alieons  could be on  the order of $10^{-28}-10^{-29}$ if they are
 uniformly distributed throughout  the earth. Or
they could be concentrated by geo-chemical processes, say, on the
surface or in the oceans,
leading to a much higher density in certain materials or regions.

 A second way to guess the abundance of ``alieons'' would be to
assume that the break in  the
spectrum at the very highest energies,  around $5\cdot10^{19} eV$,
is due to extra-galactic particles and that it is their true
incoming flux. The rate here is about one particle per $100km^2$
per year. At lower energies this component, while subdominant, will
still make some contribution. Extrapolating it to lower energies
with say a power law $E^{-2}$, we should multiply the ``knee''
estimate by $(5\cdot10^{19}/E_0) (100km^2/m^2)$ where $E_0$ is the
low energy cutoff of the extra-galactic component. If we choose 
$E_0=1GeV=10^9 eV$, this gives an enhancement factor  of $5\cdot
10^{10}\times 10^{-10}=5$. Curiously then, both guesses either on
the basis of  the ``knee'' or the ``break'', give about
the same result.

With the suggested  non-standard value of $\alpha$, the most
striking property to look for
 would evidently
be protons of non-standard charge. These would presumably end up as
molecules that are not completely neutral. The
reported~\cite{claim}  anomaly for  $\alpha$,  on the order of
$\Delta \alpha/\alpha\sim 10^{-5}$ is not extremely small by the
standards of present-day precision techniques. If
straightforwardly turned into
a change of charge, that is without concern for what is happening
to c or $\hbar$,( --not to mention other factors-- ) and assuming
negligible
``ageing'' gives a change of charge
$\Delta e/e=\frac 12 \Delta \alpha/\alpha$.  Since the
claim~\cite{claim} is for
a {\it reduced} value of  $\alpha$, a normally neutral atom or
molecule would have a slightly {\it negative} charge. Perhaps
high through-put chemical means or
molecular beam
methods could be used to concentrate and search for
 the ``alie-ocules''. By the
same token these systems (as well as their ions), although very
rare, would show distinct shifts of their spectral lines in the
laboratory and could be searched for on this basis.    

 A most dramatic aspect of  alieons would
be their escape from the exclusion principle~\cite{plaga}.
Having a
different value for some parameter such as charge, mass, or
magnetic moment,  an
``alieon'' would no longer be identical to its local counterparts,
and  the hamiltonian no longer a symmetric operator under
interchange 
of particles of the same species. This  yields a small
non-symmetric term in the hamiltonian  proportional to the
deviation of
the parameters from their conventional values. A straightforward
approach would then be to assume that for, say,  neutrons in the
nucleus, the
wavefunction for the ``alieneutron'' 
no longer need  be antisymmetric with respect to the ordinary
neutrons. 
An alien neutron captured in a nucleus would not  remain
 near the top of the neutron fermi gas of the nucleus and would
fall to the bottom of the nuclear potential, much as a hyperon in
a hypernucleus. 
 The resulting object, an ``alieotope'', would presumably be an
isotope differing in
mass from the local version by the  nuclear fermi energy, a few
tens of MeV.

Because of the  shift in nuclear energetics of the resulting
anomalous isotopes it is possible that certain  beta or electron
capture  
transitions of the nucleus, usually energetically forbidden, would
become
possible. If these could be identified, low-background experiments
of the double-beta decay type might be entertained.

A potentially experimentally interesting signal in large detectors 
could  be the observation of the energy release in gamma rays as
the alien neutron transits to the bottom of the nuclear potential
well. However, unless the degree of non-identity is very small, 
this would seem to be a rapid process and most alien neutrons
should have transited long ago. The question may be handled in a
manner analogous to that of nuclear isospin, where at first
neutrons and protons are treated as  identical and then neutron-
proton differences lead to a breaking of ``identity''. That is, the
there is a new term in the
hamiltonian, {\it antisymmetric} in the exchange of  neutron and
 alieneutron,
proportional to $(\delta p )$ the  deviation of some parameter 
such as mass or magnetic moment from its usual value. We would
then expect, very roughly, the rate of transition $\Gamma$ to be
then
$\Gamma= (\delta p/p )^2 \Gamma_{0}$ where $\Gamma_{0}$
is the typical rate of  a nuclear radiative transition,
say~\cite{ww} $\sim
10^{-14} s$. Hence unless $(\delta p/p )^2 $ is very small the
alien neutrons should transit quickly.

If there is a
 velocity-dependent ``ageing''  due to the role of proper time,as
mentioned above, we should
perhaps take into account that nucleons in the
nucleus, (or for that matter electrons in the atom or in condensed
matter) have a spectrum
of semi-relativistic velocities. If there were then ageing at
different
rates, there ought to be a gradual process of relative
``alienation'' with a slow continuous rate of Pauli principle
violating transitions.

Particles whose parameters  differ a ``little bit'' from each other
are of course
completely foreign to quantum field theory as we presently know it.
All
copies of a given type of particle should be produced by
application of  one and the same field operator. A great
revision of theory would be necessary to deal with a situation
where this is not true. Especially, the presence of different
values of the fundamental charge $e$ in the same region of space-
time would necessitate substantial changes in our present
understanding
of gauge invariance.    
These questions  and many more which quickly come to mind are
symptomatic of the fundamental difficulties,-- but perhaps also new
vistas-- that would arise  should ``alieons'' exist.



I am indebted to Harald Fritzsch  for stimulating my interest in
this
topic  and for 
 discussions,  to E.
Lorenz for a remark  on the interest of large detectors, and to D.
Semikoz for discussions on cosmic ray fluxes.




\end{document}